# How is density compensation created in the ocean mixed layer?

Andrey O. Koch, Department of Marine Science, University of Southern Mississippi, Stennis Space Center, Mississippi, USA

Robert W. Helber, James G. Richman, and Charlie N. Barron, Naval Research Laboratory, Ocean Dynamics and Prediction Branch, Stennis Space Center, Mississippi, USA

Corresponding author: A. O. Koch, Sequenom Integrated Genetics, LabCorp, 3595 John Hopkins Ct, San Diego, CA 92121, USA. (andreyokoch@gmail.com)

**Abstract**

This study examines the impact of turbulent mixing on horizontal density compensation in the upper ocean. A series of simulations model the role of mixing in scenarios initialized with geostrophically-adjusted compensated and uncompensated thermohaline gradients. Numerical experiments isolate the influence of mixing on these gradients using idealized conditions with zero surface heat and momentum flux. Prompted by theoretical considerations and observed consequences of mixing, a new non-linear horizontal diffusion scheme is introduced as an alternative to the standard Laplacian diffusion. Results suggest that when horizontal mixing is parameterized using constant diffusivities, horizontal density compensation is substantially unchanged as the gradients erode. Simulations using the new scheme, which parameterizes mixing with horizontal diffusivities scaled by squared buoyancy gradient, suggest that horizontal mixing can produce compensated gradients during this decay, but only at scales of 10 km and less. Reducing vertical mixing to small background values has a similar effect, increasing the degree of compensation at submesoscales. Reproducing observed compensated thermohaline variability in the mixed layer at scales greater than 10 km requires external forcing. These results show mixing's important influences on density compensation within the ageostrophic submesoscale regime. In the transition from a horizontally compensated mixed layer to a partially compensated thermocline, advection must play an important role; mixing alone is insufficient.

## 1. Introduction

In the upper ocean, temperature fronts are not necessarily associated with density fronts or geostrophic velocity shear. The density ($\rho$) is affected by both the temperature ($T$) and salinity ($S$), which do not have to follow a simple $T$-$S$ correlation. Horizontal density compensation in the ocean refers to conditions where horizontal gradients of temperature and salinity balance each other in their joint effect on density, substantially decreasing or canceling horizontal density gradients. The thermohaline variability in terms of potential temperature ($\theta$) and salinity effect on density is often expressed as a density ratio

$$R_\rho = \frac{\alpha\Delta\theta}{\beta\Delta S}, \qquad (1)$$

where $\alpha$ is a heat expansion coefficient, $\beta$ is a haline contraction coefficient, and $\Delta\theta$ and $\Delta S$ are the horizontal differences of potential temperature and salinity, respectively. When $R_\rho=1$, fronts exist in temperature and salinity but not in density. Thus, the horizontal density is compensated. When $R_\rho=2$, the temperature effect on density is opposite and twice as large as the salinity effect.

Observations [*Robinson*, 1976; *Robinson et al.*, 1979; *Gordon et al.*, 1982] reveal that horizontal *T,S*-gradients in the mixed layer (ML) on basin scales are uncompensated due to the differences in large-scale heat and salt forcing. *Stommel* [1993] summarizes this work and proposes a mechanism regulating this large-scale uncompensated ($R_\rho\approx2$) pattern. However, on smaller spatial scales – from tens of meters to hundreds of kilometers – density compensated *T,S*-gradients are observed everywhere in the ocean ML [*Rudnick and Ferrari*, 1999; *Rudnick and Martin*, 2002]. The following scientific questions naturally arise: (i) what are the processes and their variability scales in the upper ocean that can generate the observed horizontal density compensation? and (ii) how adequately are these processes represented in modern Ocean General Circulation Models (OGCM)?

Several processes, including atmospheric heat and freshwater fluxes, strong wind-induced vertical mixing, and internal thermohaline variability determined by advection and diffusion, are thought to be critical for density compensation to exist in the upper ocean [*Rudnick and Martin*, 2002]. In this study we are particularly interested in learning how density compensated thermohaline variability emerges from the submeso- to the mesoscales. *Fox-Kemper et al.* [2008] suggest a direct connection between the restratification of a mixed layer (slumping of a horizontal density gradient) by submesoscale mixed layer eddies and the development of compensated thermohaline gradients.

Another aspect we explore in this study is the effect of mixed layer eddies on pre-existing compensated thermohaline variability in OGCMs. While the density ratio in the ML is often close to 1, the typical density ratio in ocean thermoclines (TC) is around 2 [*Schmitt*, 1981; *Rudnick and Martin*, 2002]. Since the ocean thermocline is isolated from the direct influence of atmospheric forcing, the change in density ratio from $R_\rho\approx1$ in the ML to $R_\rho\approx2$ in the TC only can occur due to advection and diffusion. We test the ability of present model physics to reproduce this change at submeso- and mesoscales.

Theoretical works of *Chen and Young* [1995] and *Ferrari and Young* [1997] suggest the importance of mixing in developing compensated variability in horizontal thermohaline fields. If the horizontal diffusion of temperature and salinity is defined as

$$F_{\theta,S} = \nabla\big(\nu_h \nabla \cdot (\theta, S)\big), \qquad (2)$$

where $\nu_h$ is a coefficient of horizontal diffusion, then, according to abovementioned theories, creating compensated thermohaline gradients requires $\nu_h$ to be proportional to the square of buoyancy gradient:

$$\nu_h \sim (\nabla b)^2. \qquad (3)$$

We aim to assess the importance of this mixing effect on compensated variability of horizontal *T,S*-gradients in modern OGCMs by isolating it in an idealized ocean. In particular, we test the hypothesis that nonlinear horizontal diffusion in the form of equation 3 plays a critical role in generating compensated *T,S*-variability. By making our idealized ocean simulation free of external forcing, we can isolate the impact of mixing on *T,S*-variability for close evaluation in a state-of-the-art ocean circulation model that includes a complete set of ocean physics proven skillful for ocean prediction.

We perform this investigation using the Hybrid Coordinate Ocean Model (HYCOM; *Bleck* [2002]). The choice of this OGCM is arbitrary since most OGCMs share horizontal mixing schemes (equation 2) that have a constant diffusivity applied to all scalars. Our approach has two main goals: (i) constructing idealized ocean simulations capable of reproducing realistic mesoscale-submesoscale eddy fields in the upper ocean with various thermohaline initializations; and (ii) studying how the mixed layer eddy circulation and its mixing component, in particular, in the absence of external forcing, acts on pre-existing thermohaline variability in terms of density compensation.

The manuscript is organized as follows. Section 2 explains model architecture and details of different initialization settings and numerical experiments. The results of numerical simulations highlighting the impact of diffusion scaled by buoyancy gradient squared in creating compensated thermohaline variability in submesoscales are presented in section 3. Section 4 discusses the results with the emphasis on the role of mixing in compensated variability, and section 5 summarizes main results and conclusions.

## 2. Model description and experimental set-up

### 2.1 The Model and Domain

The numerical simulations in this study are performed using the HYCOM version 2.2 ocean general circulation model. The model is a free-surface primitive equation hydrostatic ocean model with generalized coordinate layers that can be depths, sigma-, or isopycnic surfaces. The coordinates adjust at every time-step allowing for simplified numerical calculation of ocean processes and providing smooth transitions from the deep ocean to coastal regimes [Bleck, 2002]. HYCOM is used for global ocean [*Chassignet et al.,* 2009] and high-resolution regional ocean prediction [e.g. Thoppil and Hogan, 2010] and model sensitivity studies [e.g. *Kara et al.,* 2010]. Algorithms that comprise a HYCOM 2.2 computational kernel are described in detail by *Wallcraft et al.* [2009].

The model domain is an *f*-plane rectangular channel centered at 40 ºN. It has 300 grid points in *x*-direction and 600 grid points in *y*-direction with horizontal resolution 0.01 degree, which gives the computational domain size of 255x665 km in *x*, *y*-directions, respectively, with horizontal spacing ~1 km. This horizontal resolution captures small scale filaments and eddies by resolving scales as small as 3km. For the model simulation the initial stratification has a Rossby radius of the order 100 km. The channel has a uniform depth, $h$=1500 m, with periodic north and south and no-slip east west boundary conditions. The net atmospheric fluxes are set to 0, meaning that all outgoing fluxes are exactly balanced by incoming fluxes. Initial conditions are solely defined by $\theta$ and $S$ fields. There are 20 hybrid vertical levels consisting of *z*-levels in the mixed layer, where isopycnals outcrop to the surface, and isopycnic levels with assigned "target" density values in the deeper regions. In HYCOM, the equation of state is a non-linear approximation of the full UNESCO equations [*Bryden et al.,* 1999] and applied such that the effect of thermobaricity [*Sun et al.*, 1999] is included in the model physics.

In the present study we will focus on the evolution of density gradients in the mixed layer, where the model coordinates are always z-levels. The vertical mixing model used, unless otherwise noted, is the K-Profile Parameterization (KPP; *Large et al.* [1994]). It effectively matches the mixing due to stronger physical processes at the upper ocean boundary layer with representations of mixing due to internal wave breaking, shear instability, and double diffusion in the ocean interior. Horizontal diffusion of tracers, unless otherwise noted, is by a horizontal Laplacian operator, dependent only on the second horizontal derivative of the diffused property with constant diffusion coefficients identical for all tracers. Lateral diffusion is separate from the vertical one-dimensional mixing of the KPP scheme. Since HYCOM has hybrid coordinates, the horizontal diffusion must account for coordinates that deviate from pressure surfaces. Using the Smolarkiewicz and Grabowski (1990) advection and diffusion scheme, the diffusion term is scaled by a neutral pressure factor that characterizes the neutral ocean state (Bleck et al., 2002), thereby separating the effects of vertical and lateral mixing and diffusion.

## 2.2 Non-linear diffusion scheme

To test the effectiveness of non-linear diffusion to enhance compensation, a new horizontal mixing scheme option is installed in HYCOM by including in equation 2 a term proportional to the scalar $\nabla b^2 = \nabla b \cdot \nabla b$ such that

$$\nu_h = \gamma \nabla b^2. \qquad (4)$$

From equation 2, horizontal diffusion of $\theta$ becomes

$$F'_\theta = \nabla \cdot (\gamma \nabla b^2 \nabla \theta) = (\gamma b_x^2 \theta_x)_x + (\gamma b_y^2 \theta_y)_y, \qquad (5)$$

and horizontal diffusion of $S$ becomes

$$F'_S = \nabla \cdot (\gamma \nabla b^2 \nabla S) = (\gamma b_x^2 S_x)_x + (\gamma b_y^2 S_y)_y, \qquad (6)$$

where the subscripts $x$ and $y$ are partial derivatives. In this new expression, $\gamma$ is a scaling coefficient estimated to give the buoyancy gradient term a similar magnitude to that of a constant $\nu_h$ case. The terms $b_x^2$ and $b_y^2$ were computed and space-time averaged in a standard simulation with constant diffusivities, and then $\gamma$ was computed such that $\nu_h = \gamma \frac{1}{2} \overline{(b_x^2 + b_y^2)}$, where $\frac{1}{2}(\overline{b_x^2} + \overline{b_y^2})$ is the average buoyancy gradient over model domain at the surface. Thus, the diffusion amount in the system when nonlinear diffusivity is used is about the same as in constant diffusivity case but it is distributed differently: the larger buoyancy gradients are the stronger is diffusion.

The buoyancy gradient squared dependence in equations (5) and (6) provide a local non-linearization of the horizontal diffusion consistent with the work of *Chen and Young* [1995] and *Ferrari and Young* [1997] but applied to a three-dimensional general ocean circulation model. This diffusion scheme will allow for diffusion dependent on the degree of density compensation. For a fixed temperature gradient, the diffusivity will decrease as the degree of density compensation increases. Thus, a compensated temperature gradient will have smaller temperature diffusion than a non-compensated gradient of the same magnitude. The same is true for salinity diffusion with the diffusion of salt decreasing as the degree of density compensation increases. As far as we know, this approach used within a 3D ocean model has not been tested before.

## 2.3 Initialization set-up

The simulations are configured to begin with geostrophically adjusted fronts with compensated or uncompensated thermohaline gradients. Our experiments will reveal how density ratios evolve during restratification. The initial model $\theta$ and $S$ fields are uniform along the channel (*y*-direction) and variable across it (*x*-direction), such that they form a horizontal density gradient in the cross-channel direction. We run two types of $\theta$ and $S$ initial fields with identical vertical stratification and cross-channel density gradient but counter-equivalent contribution of $\theta$ and $S$ (see Table 1). In the first case that represents subtropical type of front, the temperature gradient dominates and the density gradient is regarded as $\theta$-compensated with cross-channel density ratio $R_{\rho 1}$. In the second case that represents subpolar type of front, the salinity gradient dominates and the density gradient is regarded as *S*-compensated with density ratio $R_{\rho 2}$. The initial density ratios are chosen such that $R_{\rho 1} = 1/R_{\rho 2}$, which allows equal but opposite contributions of temperature and salinity gradients to the density gradient in the first and second cases, respectively. Additionally, we analyze a third case initialized by $\theta$, $S$ fields that form exactly the same cross-channel density gradient in the mixed layer as in previous two cases, but with cross-oriented horizontal $\theta$- and *S*-gradients that cooperate in their effect on density to provide density ratio $R_\rho \approx -1.0$.

These three cases provide idealized ocean scenarios that we can use to address our goals precisely. First, the initial partially density compensated gradients allow the model to reveal its capabilities in developing fully compensated thermohaline variability. Second, the existence of these density gradients is natural for the upper ocean; regions with compensated thermohaline variability tend to be interspersed by density fronts [*Rudnick and Ferrari*, 1999]. Third, the cross-channel density gradient drives along-channel geostrophic flows that will serve as a natural dynamical background for mixed layer eddies that are, in turn, also characteristic for the upper ocean.

Initial horizontal cross-channel profiles of $\theta$, *S*, and $\sigma_\theta$ in the mixed layer are shown in Figure 1. Initially, the horizontal cross-channel density profile in the mixed layer (~110 m depth) is set with maximum values of the horizontal density gradient $\partial\sigma_\theta/\partial x$ near the channel axis. The tails of the horizontal cross-channel profiles approach uniformity near the eastern and western walls, preventing upwelling and downwelling at the walls. Figure 2 shows vertical profiles of $\theta$, $S$, and $\sigma_\theta$ initial fields sampled at the western and eastern walls of the channel for the three gradient cases. Vertical density stratification is the same in each of the three cases. At the beginning of the simulation, within the ML (upper 110 m), $\theta$, $S$, and $\sigma_\theta$ fields are uniform vertically and in the *y*-direction. In the thermocline, from 110 to 150 m depth, vertical $\theta$-gradient has a range of -0.02 to -0.17 °C m$^{-1}$ in the temperature dominated gradient case (TDC; Table 1), -0.02 to -0.09 °C m$^{-1}$ in the density-cooperating gradient case (CNC), and -0.02 to -0.04 °C m$^{-1}$ in salinity dominated gradient case (SDC). Vertical *S*-gradient has a range of 0.01 to 0.02 psu m$^{-1}$ for TDC, 0.03 to 0.05 psu m$^{-1}$ for CNC, and 0.01 to 0.07 psu m$^{-1}$ for SDC; $\sigma_\theta$-gradient varies from 0.02 to 0.05 kg m$^{-4}$ in each of the three cases. From 150 m to 400 m, all properties change slightly to reach values of $\theta$=16.00 (14.85) °C, $\sigma_\theta$ =26.52 (27.29) kg m$^{-3}$ in TDC and CNC (SDC), and *S*=(36.00, 36.47, 36.66) psu in TDC, CNC, and SDC, respectively. Below the 400 m depth to the bottom, all properties have constant values.

We spin-up the model for 30 days, which is sufficient for the geostrophic flow to develop and stabilize and for vertical $\theta$ and $S$ profiles to smooth. As vertical $\theta$ and $S$ profiles are dynamically adjusted, the ML is no longer perfectly uniform in the vertical, and the cross-channel density ratio also becomes somewhat modified. To study horizontal density compensation in the ML, we use a reference depth 50 m, representative of ML, following previous studies on ML density compensation [e.g., *Ferrari and Rudnick*, 2000]. The details of the three types of initial thermohaline fields in the end of spin-up simulation are given in Table 1. By the end of the spin-up simulation, the cross-channel gradients in $\theta$, $S$, and $\sigma_\theta$ at 50 m have density ratios of $R_{\rho 1}$=4.71 and $R_{\rho 2}$=0.21 for TDC and SDC, respectively, which maintains a relationship $R_{\rho 1} = 1/R_{\rho 2}$. During spin-up simulation, the model follows only 2-dimensional (2D) dynamics in *x-z* plane because the uniform model has no source to introduce along-channel variability. In order to introduce along-channel variability that develops into an eddy field, on day 30 we perturb $\theta$ and $S$ fields with a small $O(10^{-6}$ °C, psu) random noise and run the model for additional 315 days. During that interval, submesoscale eddies develop and restratify the mixed layer; with time, they evolve into less-energetic mesoscale eddies.

The analysis of thermohaline variability and density compensation in this gradient-evolution simulation is the subject of the section 3.

### 2.4 Numerical experiment design.

Our experiments explore the evolution of density compensation under alternative representations of turbulent mixing (to be discussed in detail in section 3.4), both vertical and horizontal. To examine the impact of mixing on compensated and cooperating thermohaline variability, we run a set of numerical experiments initialized using TDC, SDC, and CNC cases of thermohaline fields in their geostrophically-adjusted states at the end of model spin-up. The different aspects of turbulent mixing we are testing include the following: a complete absence of horizontal mixing when diffusivities are set to zero; a new horizontal mixing scheme when diffusivities are a function of buoyancy gradient squared (equations 5 and 6); a case when vertical mixing coefficients are set to constant small background values, and another variation when double diffusion part of vertical mixing model is turned off.

The summary of these experiments is given in Table 2. The labels for model experiments consist of four (1—4) characters representing subsequently: (1) initialization type, (2) horizontal diffusion mode, (3) vertical diffusion mode, and (4) presence or absence of double diffusion. Each experiment will have four letter labels, where the first letter refers to the gradient type, which is T for temperature dominated, S for salinity dominated, or N for non-compensating (cooperating) gradient cases. The second letter refers to the horizontal diffusivity type with C for constant, 0 for none, or B for buoyancy gradient squared. The third letter refers to the vertical mixing model with K for KPP or B for a constant background. The fourth and last letter refers to whether double diffusion is turned on with D for on and 0 for off. For example, S0KD experiment (Table 2, 6[th] row, 5[th] column) uses a salinity-dominated gradient (SDC), zero horizontal diffusivity and KPP sub-model for vertical diffusion with double diffusion turned on.

We will describe in detail the results from two experiments that are referred to as 'basic cases' with typical diffusion characteristics. Labeled TCKD and SCKD, one uses the temperature dominated gradients and the other salinity dominated gradients. Both use constant horizontal diffusivity, and KPP with double diffusion on. We will use the basic cases as reference states in section 3.4 when considering the effects of alternate diffusion representations on mixing and evolving density compensation.

## 3. Results

### 3.1. Flow Evolution

All simulations in this numerical experiment evolve through four dynamical regimes starting with an initial geostrophic adjustment spin-up period extending to day 30 (Figure 3). The second regime is characterized by rapid growth of shallow baroclinic instability as a result of small scale noise input after spin-up on day 30 (see section 3.2). The third regime begins at approximately model day 90, when the mesoscale eddies begin to interact with the channel walls. The final regime begins near model day 135 when both kinetic energy and stratification erode as the simulations spin down.

The second regime is dominated by shallow, mixed layer baroclinic instabilities driving rapid eddy growth. The scales of the disturbances occur near the mixed layer Rossby radius of deformation given by $a=C/f$, which is approximately 15 km. Here, Rossby wave propagation speed, $C = \sqrt{gH\Delta\rho/\rho_0}$, where the mixed layer depth (110 m) sets the vertical scale, $H$, the change in density across the base of the mixed layer is $\Delta\rho \approx 2.0$ kg m$^{-3}$, the reference density is $\rho_0 \approx 1025$ kg m$^{-3}$, the Coriolis parameter is $f=0.93 \cdot 10^{-4}$ s$^{-1}$, and the gravitational acceleration is $g=9.8$ m s$^{-2}$. The Burger number is also of order one, $Bu=NH/(fa) \approx O(1)$, consistent with mixed layer instability theory [Boccaletti et al. 2007], where $N = \sqrt{-g\rho_0^{-1}\partial\rho/\partial z}$ is the buoyancy frequency. Figure 3c and 3d show the potential temperature and salinity at 50 m depth during this regime of rapid mixed layer eddy growth in the simulations. Note that in Figure 3, the panels are turned 90º so the *x*-axis becomes oriented vertically and *y*-axis horizontally.

The rapid growth of the instabilities can be seen in Figure 4 where buoyancy frequency squared (or stratification) and eddy kinetic energy are rapidly increasing early in the simulations. Eddy kinetic energy is given by $EKE = 0.5(u'^2 + v'^2)$, where $(u', v') = (u, v) - (\bar{u}, \bar{v})$ are deviations of (*x*, *y*) current components $(u, v)$ from the mean flow $(\bar{u}, \bar{v})$ and the overbars denote time-averaging. Both $N^2$ and EKE are spatially averaged over the model domain at the 50 m depth in Figure 4. The instabilities act to re-stratify the mixed layer by releasing the potential energy stored in the cross-channel density gradient. Boccaletti et al. (2007) find more rapid mixed layer instability growth rates at small scales. In Figure 5 we show the EKE separated for three length scales. At smaller length scales 3 and 10 km, we find that the growth rates are consistent with linear instability analysis for the mixed layer, which are similar to Boccaletti et al. (2007). Slopes of the EKE increase with time are consistent with the growth rates for similar size eddies shown in Figure 3 of Boccaletti et al. (2007). Notice that there are two rapid growth periods, one starting at day 50 due to rapid release of energy from rapidly growing small-scale baroclinic instabilities peaking at around model day 72, and the other near day 110 associated with the emergence of large scale baroclinic instabilities when the T and S gradients are small.

After approximately model day 90, the flow characteristics change into the third regime where the mesoscale eddies begin to interact with the channel walls (Figures 3e and 3f). During this regime, the eddy kinetic energy continues to grow but the stratification growth slows (Figure 4) as more friction at the channel walls is influencing the system. The buoyancy frequency reaches its maximum at around day 120 (Figure 4a), which is shortly followed by the peak in EKE. Looking again at the scale separated growth rates in Figure 5, we see that the slower growing larger scale deep baroclinic instabilities are emerging as the dominant instability. This again is explained in the Boccaletti et al. (2007) where the growth rates are slower at large scales. This can be seen in Figure 5, at the 100 km scale (blue line fit), representing the deep baroclinic instability growth rate.

The fourth and final regime begins near model day 135 where both stratification and eddy kinetic energy decline (Figure 4). In this regime the mesoscale eddies grow larger and smoother as the cross-channel density gradient gradually decreases. By the end of the simulation at day 345 (Figures 3f and 3h), the $\theta$ and $S$ fields become very smooth and much less energetic. In Figure 5, this is characterized as the slow decrease in EKE until the end of the simulation.

During the entire simulation, the differences in $N^2$ and EKE between the temperature and salinity dominated cases are small, suggesting that they share the same dynamical pattern in the upper ocean and the same mechanism controlling restratification. Similar to Bocalletti et al. 2007, the dynamics are governed by baroclinic instabilities only. At small 'submesoscales' mixed layer baroclinic instabilities prevail while at larger scales deep baroclinic instabilities become more dominant.

### 3.2. Time evolution of density compensation

For the purposes of characterizing the horizontal density compensation in thermohaline gradients we introduce a new measure – horizontal compensation angle (HCA) – borrowing from the concept of Turner angle (Ruddick 1983) used to characterize vertical density compensation:

$$\text{HCA} = arctan\left(\frac{R_\rho + 1}{R_\rho - 1}\right). \tag{8}$$

The horizontal compensation angle is especially useful in characterizing *T, S*-variability when fluctuations in salinity are so small that $R_\rho$ approaches infinity.

We examine the time evolution of density compensation by examining scale-dependent tendencies in $R_\rho$ and HCA. Perfectly compensated gradients have $R_\rho$=1 and HCA=±90°, indicating that the temperature and salinity contributions to density gradients perfectly balance and result in a density gradient of zero. In the TCKD case, $R_\rho$>1 and 45°< HCA<90°; a trend of decreasing density compensation would correspond to increasing $R_\rho$ and decreasing HCA. For the SCKD scenario, $R_\rho$<1 and -90°< HCA<-45°, and a trend of decreasing density compensation would be manifest as decreasing $R_\rho$ and increasing HCA.

### 3.2.1 Density compensation trajectories in the *βΔS, αΔθ* -space

In this subsection we show how little mixed layer density compensation changes over the course of cross-channel density gradient evolution. In Figure 6, we show the density compensation evolution at 50 m depth representing density ratio and horizontal compensation angle as a function of $\beta\Delta S$ and $\alpha\Delta\theta$, thus displaying density compensation in a density coordinate space [cf. *Ruddick,* 1983]. If we consider a point on a compensation trajectory, the radius drawn from the coordinate origin through the point marks the $R_\rho$ and HCA values on the circumference, while corresponding points on the *x* and *y* axes represent values of $\beta\Delta S$ and $\alpha\Delta\theta$. We focus on temperature and salinity differences at three spatial scales, 3, 10, and 100 km. The 3 and 10 km length scales characterize the submesoscale, while the 100 km length scale characterizes the mesoscale. Every 12 hours at 50 m, we compute medians of differences $\alpha\Delta\theta$ and $\beta\Delta S$ over each spatial scale in a cross-channel section which are then averaged in the *y*-direction to form the time series. Figure 6, therefore, shows the relative contribution of salinity and potential temperature to the density gradients at the 3, 10, and 100 km spatial scales giving a consistent representation of thermohaline structure. The general structure of the trajectories is similar for all scales. All the trajectories follow nearly uniform radial heading towards origin with zero gradients consistent with nearly constant HCA and $R_\rho$. The times of the dynamical model states depicted on Figure 3 are marked on the trajectories to show the connection between dynamics and density compensation.

The gradient magnitude is decreasing towards the origin more rapidly at the 3 and 10 km scales than at the 100 km scale (Figure 6). This can be seen by noting that day 72 is closest to the origin in Figure 6a (3 km scale) and slightly further away from the origin in Figure 6b (10 km scale). At the 100 km scale, day 72 has hardly moved from the initial gradient magnitude.

At the 3 and 10 km scales, submesoscale stirring [e.g., *Klein et al.,* 1998] causes density compensation to increase from day 30 until day 70 - 80, the time dominated by mixed layer instability growth. The small-scale $\theta$, *S*-perturbations around day 48 reflect this initial increase in density compensation as a result of the rapid mixed layer eddy growth discussed earlier and reflected on day 72 (Figures 3c and 3d). The phase of increased density compensation at submesoscales (Figures 6a and 6b) corresponds to the phase of the EKE increase (Figures 4b and 5) that, again, shows a direct effect of mixed layer instabilities on compensated thermohaline variability. The trajectory wiggles found only at smaller scales (Figures 6a and 6b) and confined between days 70-100 represent the emergence of the deep baroclinic instability which is reflected in somewhat reduced density compensation and an increase in density gradient due to mesoscale stirring. This can be seen in Figures 6a and 6b near day 72 for both TCKD and SCKD, where the red and blue lines move toward the 45° and -45° HCA, respectively. Since in Figure 6 density iso-lines are parallel to the diagonal passing through the -90° and 90° HCA points, trajectories towards the upper left quadrant correspond to increasing density.

After this emergence of stirring, density compensation, again, increases reaching the earlier level which corresponds to the second growth of mixed layer instabilities (Figure 5). Further on, the thermohaline trajectory descends to the coordinate origin which corresponds to the first 0 crossing by $\Delta\sigma_\theta$ on day 153 for the TCKD and on day 174 for the SCKD experiencing just a slight decrease in density compensation. From that point until the end of the simulation, the density compensation trajectory fluctuates about the coordinate origin during which time the gradients succumb to turbulent diffusion and density compensation continues to drop slightly.

Summarizing this subsection, we can say that density compensation trajectories do not change much in the course of mixed layer evolution, temperature- and salinity-controlled cases are very alike, and initial increase in density compensation associated with mixed layer baroclinic instability is present only at smaller scales (3, 10 km). They start at around $R_\rho$=4 (0.2) in TCKD (SCKD) and descend with time to the coordinate origin where they fluctuate about it. At the 100 km scale, slower deep baroclinic instability dominates and the evolution towards the coordinate origin is slower.

### 3.2.2 Transition between mixed layer and thermocline types of density compensation

In this subsection we simulate a conceptual change in compensation from mixed layer to thermocline and show that it is insubstantial compared with the observations. For example, it has been observed in the ocean that $R_\rho$ changes from 1 in the ML to 2 in the TC [e.g., *Rudnick and Martin,* 2002]. In order to compare the present idealized case of density compensation evolution with the observed case, we plot on Figure 6 the sectors representing $R_\rho$ confined in the interval (1, 2) or HCA=(71.5º, 90º) for both $\theta$- and $S$-dominated regions. The area taken by the sector represents the transition in the $\alpha\Delta\theta$, $\beta\Delta S$-space between two different compensation states. In contrast, the model density compensation trajectories on Figure 6 tend to remain within a narrow HCA diapason, representing the same state of density compensation throughout the whole simulation period without any selectivity for creating density compensation as suggested by observations. To support this visual impression, for HCA computed from $\alpha\Delta\theta$, $\beta\Delta S$ and presented on Figure 6, we compute standard deviations (STD) and root mean square errors, $\text{RMSE} = \sqrt{\overline{(\text{HCA} - \text{HCA}_{30})^2}}$, where HCA$_{30}$ is the HCA initial value on day 30 and the overbar denotes ensemble mean. The values of both RMSE and STD presented in Table 3 are small and vary insignificantly for all spatial scales and for both basic experiments suggesting low variance and a persistence of the initial compensation state. To make a more proper quantitative comparison between the model and the real ocean, we estimate HCA ranges (RA) taking 3*STD as a range criterion. The range is estimated as a difference between maximum and minimum of the ensemble, but since we do not have actual observations, the ±3*STD=6*STD estimate is used instead. Six standard deviations for the Gaussian distribution represent a range with the 98% confidence level. For the modeling results, we have RA=6*STD, since the compensation fluctuates about the mean of a single compensated state. For the observations, the range is RA=90+3*STD-(71.5-3*STD) since there are two compensation states (71.5º, 90º) that the ±3*STD criterion is applied to. The RA estimates for both modeling results and observations are summarized in Table 3. In most instances, the ranges reconstructed for the ML-TC density compensation change scenario are 2 to 3 times larger than the corresponding ranges computed from the model. Moreover, this difference is expected to be even greater if, instead of modeled, the observed HCA STDs were taken for the ranges reconstruction. Therefore, the advection and diffusion mechanisms in the present model have limited ability in changing the initial state of horizontal density compensation compared to the real ocean.

### 3.2.3 Scale-dependence of density compensation

In this subsection we explore how density compensation depends on spatial scales. To demonstrate this scale dependence, $R_\rho$ is averaged over the 315 days simulation and plotted against spatial scales in Figure 7 to summarize density compensation in the cross-gradient direction at all spatial scales. At larger scales, in both the temperature- (Figure 7a) and salinity-dominated (Figure 7b) cases temperature and salinity variability barely differs from the initial level of compensation. At small scales, stirring produced by the rapidly growing instabilities sharpens *T,S*-gradients and enhances compensation. Therefore, density gradients at submesoscales tend to be more compensated than at mesoscales. For the TCKD, this differentiation between smaller (3-10 km) and larger (>20 km) scales (Figure 7a) is more pronounced than for the SCKD (Figure 7b). The $R_\rho$ goes unstable as the scale approaches values close to the basin width, likely because of reduced estimate robustness due to low sample volume.

More statistical properties of density compensation at different spatial scales can be revealed by probability density function (PDF) of horizontal compensation angle. The PDFs of HCA for TCKD and SCKD cases at $\Delta x$=(3, 100) km computed for the active mixed layer instability phase and the rest of model integration period are presented in Figure 8. The PDF mode stays at |HCA|=55º, $R_\rho$=5.7 (0.18) for both spatial scales and for both stages of model integration in experiments TCKD (SCKD). These numbers are very close to the initial compensation level for all spatial scales (Figure 6) which, again, shows a limited model skill in changing the density compensation state. At 3 km spatial scale during active mixed layer instability growth phase, the eddy stirring broadens and flattens the HCA PDF comprising larger |HCA| values, i.e. stronger compensation (Figures 8a and 8e), which agrees with Figure 7, and in the TCKD (Figure 8a) this tendency is more pronounced suggesting existence of more small-scale compensated density gradients than in the SCKD. However, the PDF mode is not shifted and stays at the same rate |HCA|=55º. It is interesting to note that at 3 km scale there is a small increase in PDF in the opposite end of HCA spectrum, i.e. at negative HCA values for TCKD and at positive HCA values for SCKD. This can be explained by submesoscale eddies stirring $\theta$ and *S* fields in a way that in some instances |$\beta\Delta S$| slightly exceeds |$\alpha\Delta\theta$| for TCKD and vice versa for SCKD. At 100 km scale, the HCA PDF shape is almost not changed in a transition from active eddy phase to final less energetic stage of the model integration suggesting that large-scale gradients are not sensitive to compensation-facilitating eddy dynamics.

Although there are some differences in the evolution of horizontal density compensation between TCKD and SCKD, the model shows the same general behavior of horizontal density compensation in both cases. While the system passes through different dynamical stages of thermohaline gradients evolution, the model physics does not allow density compensation to change substantially in both numerical settings. The similarity in density compensation evolution between TCKD and SCKD experiments best shown in Figures 7 and 9 reflects the similar dynamical behavior between these experiments (Figure 4). A limited model ability to produce or change density compensation state along with similarities in its evolution for $\theta$- and *S*-dominated cases leads us to investigate the details of diffusion parameterization used in the model. The next section explores the effects of model diffusion parameterization on density compensation.

### 3.3. Effect of mixing on density compensation
### 3.3.1 Effect of horizontal mixing on density compensation

In this subsection we explore effect of different implementations of horizontal mixing on density compensation. We analyze experiments with constant horizontal diffusivity (TCKD) and zero diffusivity (T0KD, Table 2). Also, we analyze the experiment TBKD (Table 2) when temperature-dominated initial gradient case (TDC) is run with variable diffusivities dependent on buoyancy gradient squared (equations 5 and 6). We do this by showing the data in two formats.

Figure 9 in T/S density scaled gradient space and Figure 10 where the density compensation is displayed as a function of scale. The lowest order characteristics of the experiments are similar, we are exploiting the higher order differences which govern the evolution of density compensation.

Following *Chen and Young* [1995] who also used nonlinear diffusivities in form of equation (4) in their analytical model, we use relative contributions of salinity $\beta\Delta S$ and temperature $\alpha\Delta\theta$ as a measure of density compensation for experiments with constant (TCKD), zero (T0KD), and nonlinear (TBKD) horizontal diffusivities. The joint distribution function, for a range of spatial scales in the $\beta\Delta S$ and $\alpha\Delta\theta$ space at 50 m depth, is given by $F(\beta\Delta S, \alpha\Delta\theta)$. In Figure 9, the natural logarithm of $F(\beta\Delta S, \alpha\Delta\theta)$ is plotted for cases TCKD, T0KD, and TBKD. We also plot the log of the absolute value of the differences of $F(\beta\Delta S, \alpha\Delta\theta)$ for the T0KD and TBKD versus TCKD cases times the sign of their differences (Figures 9c and 9f). Figure 9 is for model day 75, at the peak of active mixed layer instability growth when mixing is expected to have strongest effect. In the constant diffusivity case (Figure 9a), the cloud of possible thermohaline gradients is separated in two elongated yellow branches corresponding to larger-scale, less compensated gradients established by the initial cross-channel *T,S*-structure and smaller-scale, more compensated gradients induced by mixed layer submesoscale eddies. The black and magenta lines present in all plots of Figure 9 corresponding to initial large-scale density ratio $R_\rho$=4.7 (Table 2) and TCKD small-scale ratio $R_\rho$=2.2 (Figure 10, left end of solid line) align with the two elongated yellow branches on Figure 9a. Turning off diffusion (Figure 9b) allows stirring to disperse thermohaline characteristics over larger ($\beta\Delta S$, $\alpha\Delta\theta$) space than in constant diffusivity case, thus increasing the number of different possible thermohaline gradients and more compensated in particular. The two-branch structure of the thermohaline cloud is also present in T0KD, though it is not as prominent as in TCKD. The second branch, representing small-scale variability, turns towards the compensation line, reflecting how stirring facilitates density compensation in non-diffusive ocean. In Figure 9c, representing difference between T0KD and TCKD, this effect is shown by the red branch on the right tending to align with small-scale density ratio ($R_\rho$=1.3) yellow line (Figure 10, left end of dashed line). When diffusivity is scaled by the buoyancy gradient squared (Figure 9e), its effect on density compensated gradients resembles that of TCKD with an important difference that the right branch of the probability distribution cloud turned closer towards compensation line like in T0KD. However, this branch is more prominent than in T0KD. In Figure 9f the red branch is aligned with small-scale density ratio ($R_\rho$=1.6) yellow line (Figure 10, left end of dot-dashed line).

The differences between cases with different diffusion implementation seen only marginally in Figure 9 are better revealed in Figure 10. On Figure 10 we plot spatially- and time-averaged horizontal cross-channel density ratio $R_\rho$ as a function of a spatial scale $\Delta x$ for experiments TCKD, T0KD, and TBKD (see Table 2) at 50 m depth. One can see scale separation for all experiments when compensation increases at small scales and decreases at large scales. Setting horizontal diffusivities to zero (T0KD) has strongest effect on density compensation dropping $R_\rho$ as low as 1.3 at smallest permitted scales (Figure 10, dashed line). Introducing diffusivities dependent on buoyancy gradient squared (TBKD) also increased density compensation setting $R_\rho$=1.6 at smallest scales (Figure 10, dot-dashed line). It is of a big importance to note that changing diffusivities from constant to zero and nonlinear has an appreciable effect on density compensation only at scales smaller than 20 km.

Thus, we have shown that the nonlinear diffusion described by the theory (equation 3) makes thermohaline variability more compensated especially at smaller scales. This result suggests scale-selectivity of the stirring effect on density compensation. At submesoscales, the thermohaline gradients are sharper and thus enhance density compensation through stirring more than at mesoscales. However, results for TCKD, T0KD, and TBKD do not differ tremendously. In Figure 10 we see that smaller scales are separated from larger scales for all three experiments though in T0KD and TBKD compensation is increased. Next we test the effect of mixing parameterization on a cooperative gradient where temperature and salinity both contribute to the density gradient.

### 3.3.2 Effects of vertical, horizontal mixing, and their combination on density compensation

In this subsection we explore a fundamentally different initialization setting, where temperature and salinity gradients both act on density with the same sign and magnitude, the cooperating (non-compensating) density gradient regime (CNC, Table 1). In cooperating density gradient setting, *T, S*-gradients tend to align with cooperation line (with $R_\rho$=-1) bisecting upper left and lower right quadrants of the ($\beta \Delta S$, $\alpha \Delta \theta$) coordinate space (Figure 11). The CNC regime contrasts the compensated temperature- or salinity-dominated settings (TDC and SDC), where *T,S*-gradients tend to align with compensation line (with $R_\rho$=1) bisecting upper right and lower left quadrants (Figure 9), Because the CNC regime follows the cooperation line, as opposed to the compensation line, these gradients have the opposite meaning.

To evaluate the effects that vertical and horizontal mixing and their combined action have on density compensation we use a new case where the KPP vertical mixing sub-model is turned off, leaving only small constant vertical diffusivities for vertical mixing (Table 2; cases NCB0 and NBB0). With this new model configuration in the CNC initialization setting we run 4 experiments alternating between constant and nonlinear horizontal diffusivities and between KPP-defined and constant vertical diffusivities (see Table 2). To contrast and compare the effects of these different diffusion settings on density compensation, in Figure 11 we plot a function of joint distribution of $\beta \Delta S$ and $\alpha \Delta \theta$ in the same manner as in Figure 9. We look at the range of time instances and in Figure 11 show a time instant during the peak of mixed layer instability growth. For the cooperating gradient initialization (CNC) with constant horizontal diffusivities and KPP vertical mixing model (experiment NCKD, see Table 2), the scatter-plot cloud of ($\beta \Delta S$, $\alpha \Delta \theta$) reflects cooperating effect of temperature and salinity gradients in the elongated form and the orientation of the cloud along cooperation line with $R_\rho$=-1. Notably, having visual deviation from NCKD in shape and size of the scatter-plot cloud for other experiments (NBKD, NCB0, and NBB0), the most important feature characterizing density compensation rate – the cloud orientation – remains almost unaffected and still aligned with the cooperation line. It means that neither of these three innovations – density cooperating *θ,S*-gradients, nonlinear horizontal mixing, and small constant vertical mixing – in our model setting are able to move cooperating thermohaline gradients out of upper left and lower right quadrants. The small differences in the ($\beta \Delta S$, $\alpha \Delta \theta$) clouds for different experiments seen in Figure 11 are not informative enough and lead us to explore them in a different representation, when density compensation is displayed as a function of scale, similarly to the previous section.

To demonstrate the actual significance of the non-linear diffusion scheme we use the scale-selective representation of density compensation. The density ratio as a function of spatial scales computed in the same manner as in Figure 10 is shown in Figure 12 for the experiments NCKD, NBKD, NCB0, and NBB0. While the curve for the basic experiment NCKD with constant horizontal and KPP-defined vertical mixing is close to most uncompensated state $R_\rho$=-1, the curves for other experiments deviate from NCKD towards more compensated region at spatial scales less than 10 km. While the case NCB0, when only vertical diffusivities were changed to small constant values, only slightly deviates from NCKD, the other two cases NBKD with nonlinear horizontal mixing and NBB0 with nonlinear horizontal and constant vertical mixing show much greater effect on density compensation. The largest increase in density compensation happens in case NBB0 likely due to a combined effect of introducing nonlinear horizontal mixing and almost eliminating vertical mixing. This is one of key results of this research, which shows how horizontal mixing can change density compensation at submesoscales and that the effect is mitigated by vertical mixing.

It is also helpful to note the importance of shallow small-scale baroclinic instabilities in modifying density compensation. As an addition experiment, we estimated EKE growth rate at 300 m (not shown) and discovered that it was much slower than at 50 m (Figure 5), which is consistent with *Capet et al.* [2008]. At the smaller scales in shallow mixed layer instabilities grow rapidly and provide more energy to take advantage of the non-linear advection scheme and facilitate density-compensated gradients. At depth, this effect is absent.

## 4. Discussion

Results suggest the initial state of horizontal mixed layer compensation can be changed by nonlinear diffusion dependent on the buoyancy gradient squared but only at submescales. *Fox-Kemper and Ferrari* [2008] consider a similar dynamical situation to the one presented in this work. They set up a realistic submesoscale-resolved 3D ocean model, which produces mixed layer eddies, along with a 2D coarse prognostic model, where mixed layer eddies are parameterized. The 3D model produces a buoyancy flux which is underrepresented in the 2D model but proportional to the flux in the 3D model. The authors argue that the 2D model misses the diabatic across-isopycnal part of buoyancy flux, or residual flux, which is responsible for front slumping and widening. This residual flux could be reproduced by either multiplying the 2D buoyancy flux by a constant or by increasing the horizontal diffusion. Therefore, increased diffusivities reproduce the unrepresented part of submesoscale eddy fluxes facilitating density front slumping and mixed layer restratification. However, *T* and *S* mixing with constant diffusivities could prevent development of the compensated thermohaline variability which is evident from our results (e.g., Figure 6).

The attempts to implement more sophisticated horizontal diffusion schemes in ocean models have been made in idealized 1D and 2D frameworks. The distinguishing feature of these models is a nonlinear dependence of the diffusion on horizontal buoyancy gradient [cf. *Chen and Young,* 1995; *Ferrari and Young,* 1997]. *Ferrari and Young* [1997] establish a thoroughly-mixed upper layer with randomly distributed *T* and *S* in their model and spin-down it with *T,S*-diffusivities dependent on horizontal buoyancy gradient squared. The idealized one-dimensional model developed strongly compensated gradients due to the joint effect of the nonlinear diffusion of the thermohaline gradients on buoyancy. When we implement the same buoyancy-gradient squared horizontal mixing scheme in our three-dimensional ocean circulation numerical model, the thermohaline variability increases the compensation of density gradients at submesoscales.

The processes of vertical and horizontal turbulent diffusion in the upper ocean are interconnected. *Smith and Ferrari* [2009] show competing roles of vertical and horizontal turbulent diffusions in the upper ocean for tracer redistribution. The sensitivity of our results to the details of vertical and horizontal diffusion (Figure 12), especially the fact that horizontal diffusion becomes important on smaller scales where vertical motions are comparable to horizontal motions, also suggests a similar effect on density compensation. Therefore, a mixing scheme where vertical and horizontal diffusions are coupled could be a plausible alternative to existing mixing schemes where vertical and horizontal diffusions are independent.

We postulate that the limited ability of the OGCM used in this study to generate and change density compensation is partially due to the horizontal diffusion scheme which maintains the constant diffusivities regardless of the spatial scale and treats all tracers including temperature and salinity in the same fashion. Using spatial scale- and variable-independent diffusivities is common for existing operational OGCMs. The nonlinear horizontal mixing scheme, which is not commonly used, helped to increase compensated $\theta,S$-variability, but its effect was confined to the submesoscales smaller than approximately 10 km (Figures 10, 12). In order to improve the OGCM's ability of generating compensated thermohaline variability appropriately at larger spatial scales, as is generally observed in the ocean, it is necessary to force the model externally and generate stirring facilitating density compensation over a range of spatial scales. Mixing alone does not generate the compensated density gradients on larger scales observed in the ocean.

## 5. Summary and conclusions

Based on the state-of-the-art 3D operational oceanic general circulation model HYCOM, we built an idealized periodic channel simulation to explore the ability to develop horizontal density compensation in the upper ocean and change the initial temperature and salinity structure. We studied three basic scenarios for the initial cross-channel density gradient in the mixed layer, where it is due to (i) a temperature gradient, (ii) a salinity gradient, and (iii) cooperating temperature and salinity gradients equally contributing to the density gradient. In the first two scenarios, density compensation decreased slightly with time. The process of decreasing horizontal density compensation is consistent with the restratification of the mixed layer by submesoscale eddies that cause density and thermohaline gradients to decay.

Despite some differences in density compensation evolution for temperature- and salinity-controlled settings, for both cases, the model physics was unable to substantially change initial state of horizontal density compensation (Figure 6) towards the compensated state suggested by observations [e.g., *Rudnick and Martin,* 2002]. During the evolution of the horizontal thermohaline gradient, stirring broadened the distribution of density compensation (Figure 8), but did not shift the initial degree of compensation. Density compensation in the model is not selectively enhanced in this run-down experiment as observations of the ocean suggest must occur. Since the ocean has the capability to enhance density compensation dynamically, our model must be missing the physics required for density compensation development.

Based on the theoretical work of *Chen and Young* [1995] and *Ferrari and Young* [1997], we added new physics to our model simulations by implementing of a nonlinear horizontal mixing scheme with diffusivities scaled by buoyancy gradient squared (equation 3). Simulations with the new horizontal mixing scheme partially changed the distribution of compensated thermohaline gradients. Density compensation became enhanced at submesoscales. These results show the importance of using a nonlinear diffusion scheme in operational OGCMs to facilitate compensated variability through mixing.

Density gradients in the submesoscale tend to be more compensated than in the mesoscale. The density compensation dependence on spatial scale is more pronounced when submesoscale eddies are at their most active phase. Mixed layer eddies are particularly important in creating compensated thermohaline variability.

The sensitivity analysis showed clear dependency of the horizontal density compensation on diffusion parameterization. When an initial density gradient was formed by equally contributing cooperating $\theta$- and $S$-gradients, the implementation of nonlinear horizontal mixing (equations 5 and 6) also enhanced density compensation. In this setting the scale-dependence of compensated gradients development was more pronounced: only scales less than 10 km were susceptible to an increase in density compensation. Moreover, setting vertical diffusivities to small constant background values in a combination with nonlinear horizontal mixing scheme has further increased density compensation at submesoscales.

These findings suggest a need for the incorporation of a horizontal mixing scheme with diffusivities scaled by buoyancy gradient squared in operational OGCMs for more realistic reproduction of thermohaline variability in the upper ocean. Further development of coupled horizontal and vertical mixing scheme capable of differentiating between variability scales as well as between tracers and implementation of that scheme in operational OGCMs is another approach to this problem.

Although nonlinear horizontal mixing is capable of enhancing compensated variability in the upper ocean in the absence of external forcing, its effect is solely confined to submesoscales. In order to reproduce compensated thermohaline variability in a most realistic way over a range of spatial scales it is essential to provide external forcing to the system.

**Acknowledgements**

The Naval Research Laboratory contribution to this research is funded by the Office of Naval Research program element 61153N The Impact of Spice on Ocean Circulation. Discussions with A. Wallcraft helped considerably in optimizing our model configuration. Suggestions from P. Thoppil and X. Xu on initial model set-up were also helpful.

| Gradient type | Label | $R_\rho$ | HCA* | $\Delta\sigma_x$, kg m$^{-3}$ | $\Delta\theta_x$, °C | $\Delta S_x$, psu |
|---|---|---|---|---|---|---|
| Temperature dominated gradient, compensated | TDC | 4.71 | 57 | -1.25 | 6.03 | 0.49 |
| Salinity dominated gradient, compensated | SDC | 0.21 | -57 | -1.23 | -1.38 | -2.01 |
| Cooperating density gradient, non-compensated | CNC | -1.03 | 0.8 | -1.33 | 2.82 | -0.81 |

Table 1. A summary of the horizontal characteristics of the three gradient types after initialization and 30 day spin-up: density ratio $R_\rho$, horizontal compensation angle HCA (* defined in section 3.3), across-channel mixed layer differences in density $\Delta\sigma_x$, temperature $\Delta\theta_x$, and salinity $\Delta S_x$.

| Gradient type | Diffusion Parameterization | | | Simulation Case Label |
|---|---|---|---|---|
| | Horizontal | Vertical | Double | |
| Temperature dominated gradient, compensated, TDC | Constant diffusivity, $\nu = 5.0\ m^2 s^{-1}$ | KPP | ON | TCKD |
| | Zero diffusivity, $\nu = 0.0\ m^2 s^{-1}$ | KPP | ON | T0KD |
| | Constant diffusivity, $\nu = 5.0\ m^2 s^{-1}$ | KPP | OFF | TCK0 |
| | Buoyancy gradient squared, $\nu_h = \gamma \nabla b^2 *$ | KPP | ON | TBKD |
| Salinity dominated gradient, compensated, SDC | Constant diffusivity, $\nu = 5.0\ m^2 s^{-1}$ | KPP | ON | SCKD |
| | Zero diffusivity, $\nu = 0.0\ m^2 s^{-1}$ | KPP | ON | S0KD |
| | Constant diffusivity, $\nu = 5.0\ m^2 s^{-1}$ | KPP | OFF | SCK0 |
| Cooperating density gradient, non-compensated, CNC | Constant diffusivity, $\nu = 5.0\ m^2 s^{-1}$ | KPP | ON | NCKD |
| | Buoyancy gradient squared, $\nu_h = \gamma \nabla b^2 *$ | KPP | ON | NBKD |
| | Constant diffusivity, $\nu = 5.0\ m^2 s^{-1}$ | Background, $\nu_z = 0.004\ m^2 s^{-1}$ | OFF | NCB0 |
| | Buoyancy gradient squared, $\nu_h = \gamma \nabla b^2 *$ | Background, $\nu_z = 0.004\ m^2 s^{-1}$ | OFF | NBB0 |

Table 2. A summary of the idealized simulation cases defined by their gradient type and horizontal, vertical, and double diffusion settings. Gradient type is characterized by the initial across channel characteristics. Table 1 contains details about the frontal characteristics of each case. *The constant $\gamma$ is chosen to satisfy $\nu_h \approx \gamma \frac{1}{2} \overline{(b_x^2 + b_y^2)}$.

| factor | STD | | | RMS | | | RA | | |
|---|---|---|---|---|---|---|---|---|---|
| exp. | 3 km | 10 km | 100 km | 3 km | 10 km | 100 km | 3 km | 10 km | |
| model | TCKD | 3.9 | 5.9 | 7.1 | 4.0 | 5.8 | 8.2 | 23.4 | 35.4 |
| model | SCKD | 3.1 | 3.2 | 1.6 | 4.8 | 6.3 | 2.9 | 18.6 | 19.2 |
| ocean | TCKD | - | - | - | - | - | - | 41.9 | 53.9 |
| ocean | SCKD | - | - | - | - | - | - | 37.1 | 37.7 |

Table 3. Standard deviations (STD), root mean square errors (RMSE), and ranges (RA) of horizontal compensation angle (HCA, º) for 3, 10, and 100 km spatial scale in TCKD and SCKD from model simulations and from conceptual estimation of the real ocean scenario when $R_\rho$ changes in the range (1,2).

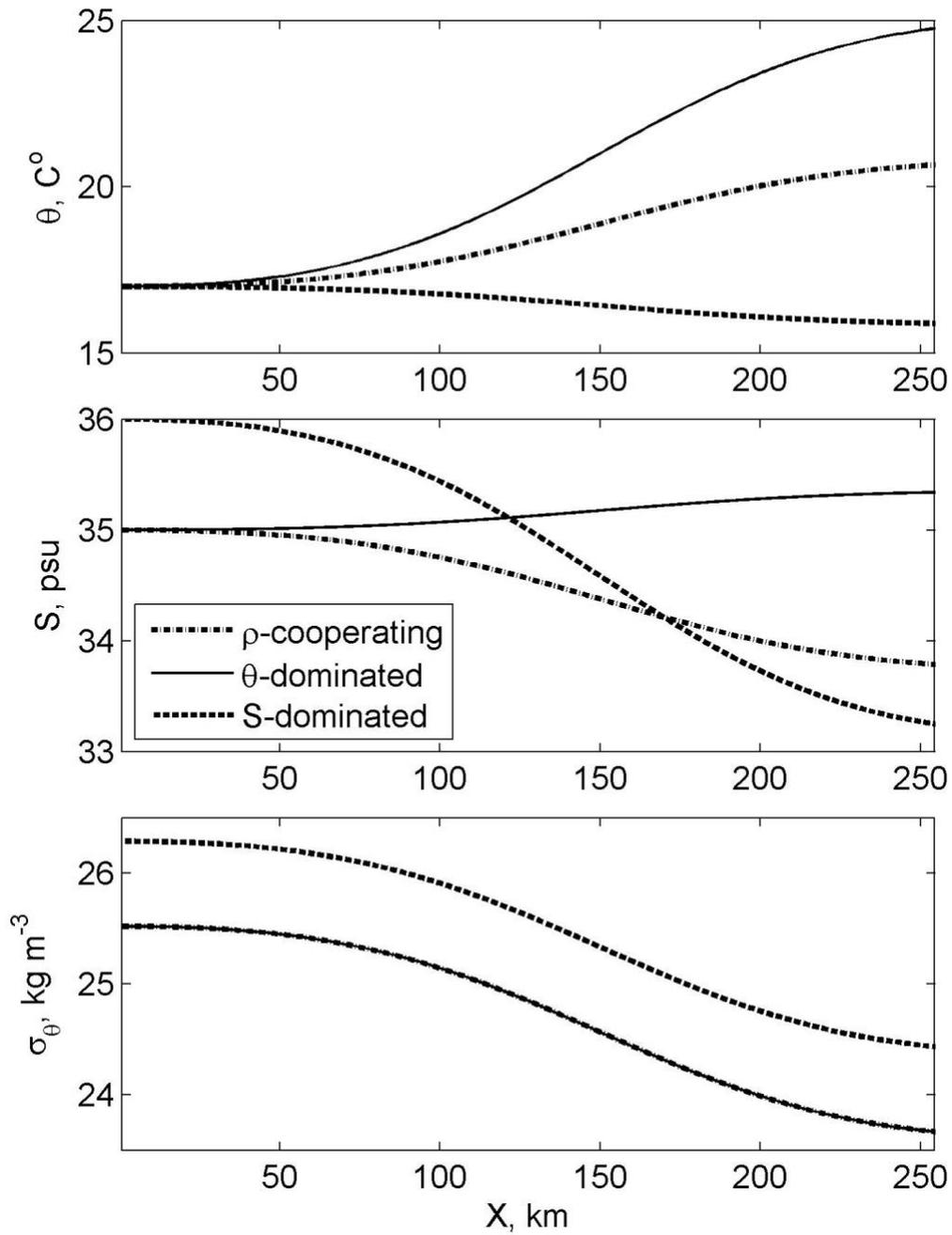

Figure 1. Initial cross-channel mixed layer profiles of potential temperature ($\theta$, a), salinity ($S$, b) and potential density ($\sigma_\theta$, c) for (solid line) temperature-dominated, (dashed line) salinity-dominated, and (dash-dotted line) density-cooperating cases at 50 m depth (see Table 1).

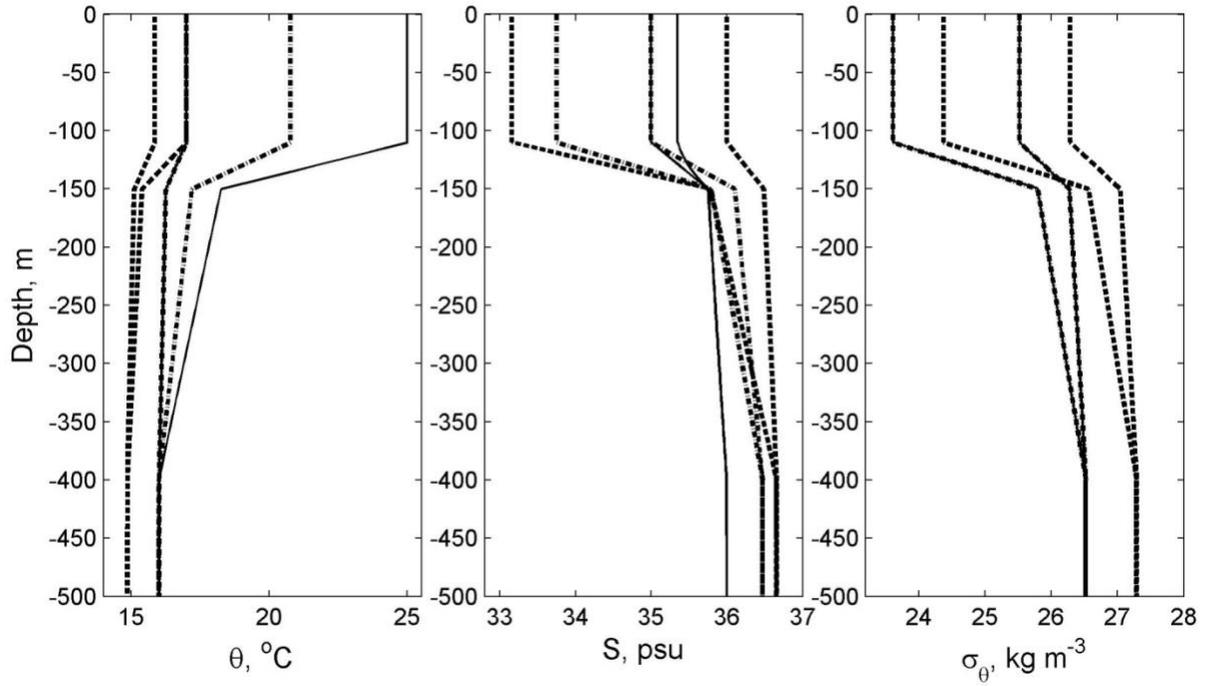

Figure 2. Initial vertical profiles of $\theta$, $S$, and $\sigma_\theta$ sampled at the western and eastern walls of the channel for (solid line) temperature-, (dashed line) salinity-dominated, and (dash-dotted line) density-cooperating cases (see Figure 1).

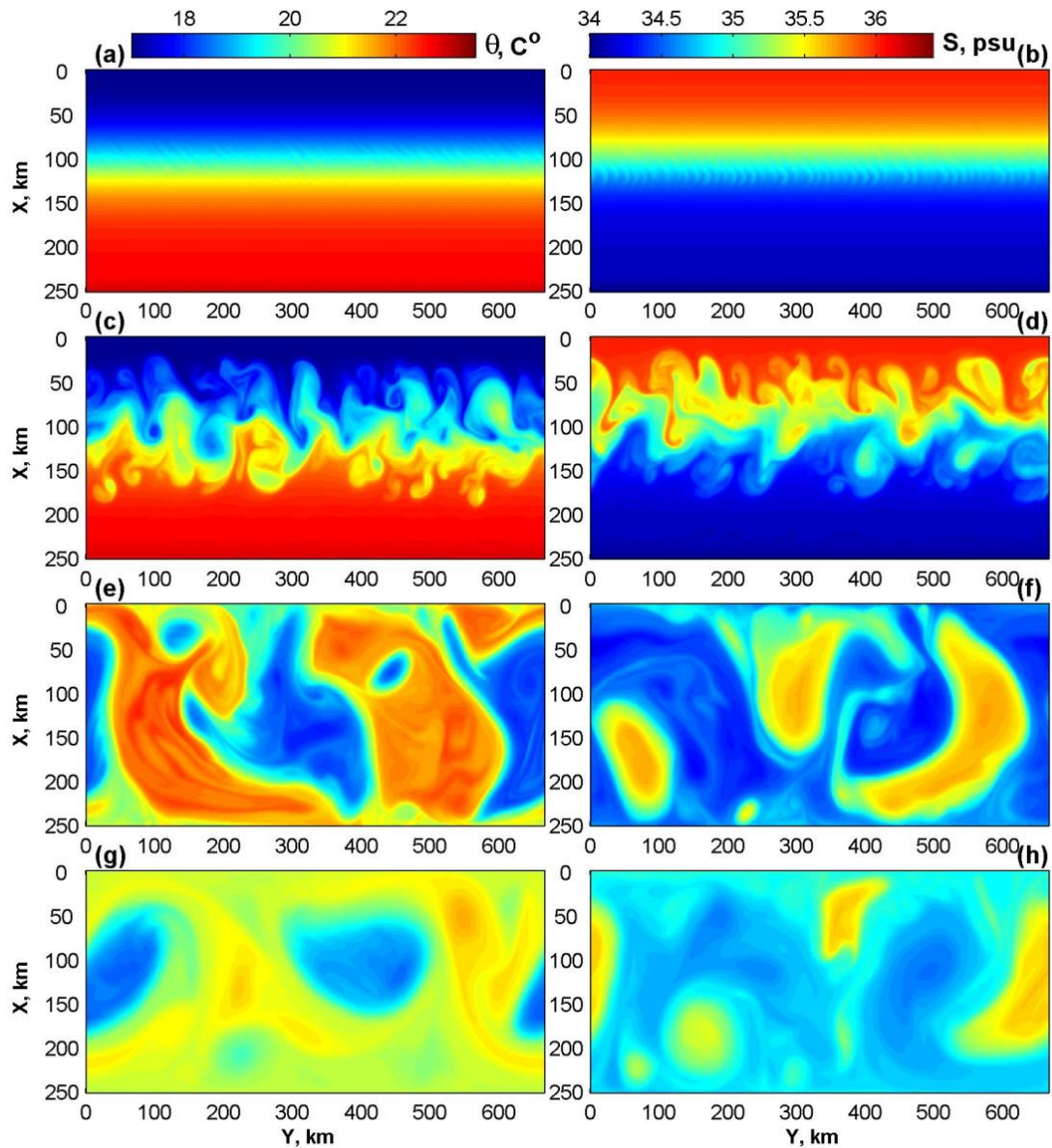

Figure 3. Horizontal planes of (a, c, e, g) potential temperature ($\theta$, ºC) for TCKD and (b, d, f, h) salinity (*S*, psu) for SCKD at 50 m depth. The plates show different stages of mixed layer evolution: generation of along-channel variability on day 48 (a, b), vigorous submesoscale variability on day 72 (c, d), energetic mesoscale eddies on days 153 (e) and 174 (f), smoothed and less energetic field in the end of the simulation on day 345 (g, h). The plates are turned 90º so the *x*-axis becomes oriented vertically and *y*-axis horizontally.

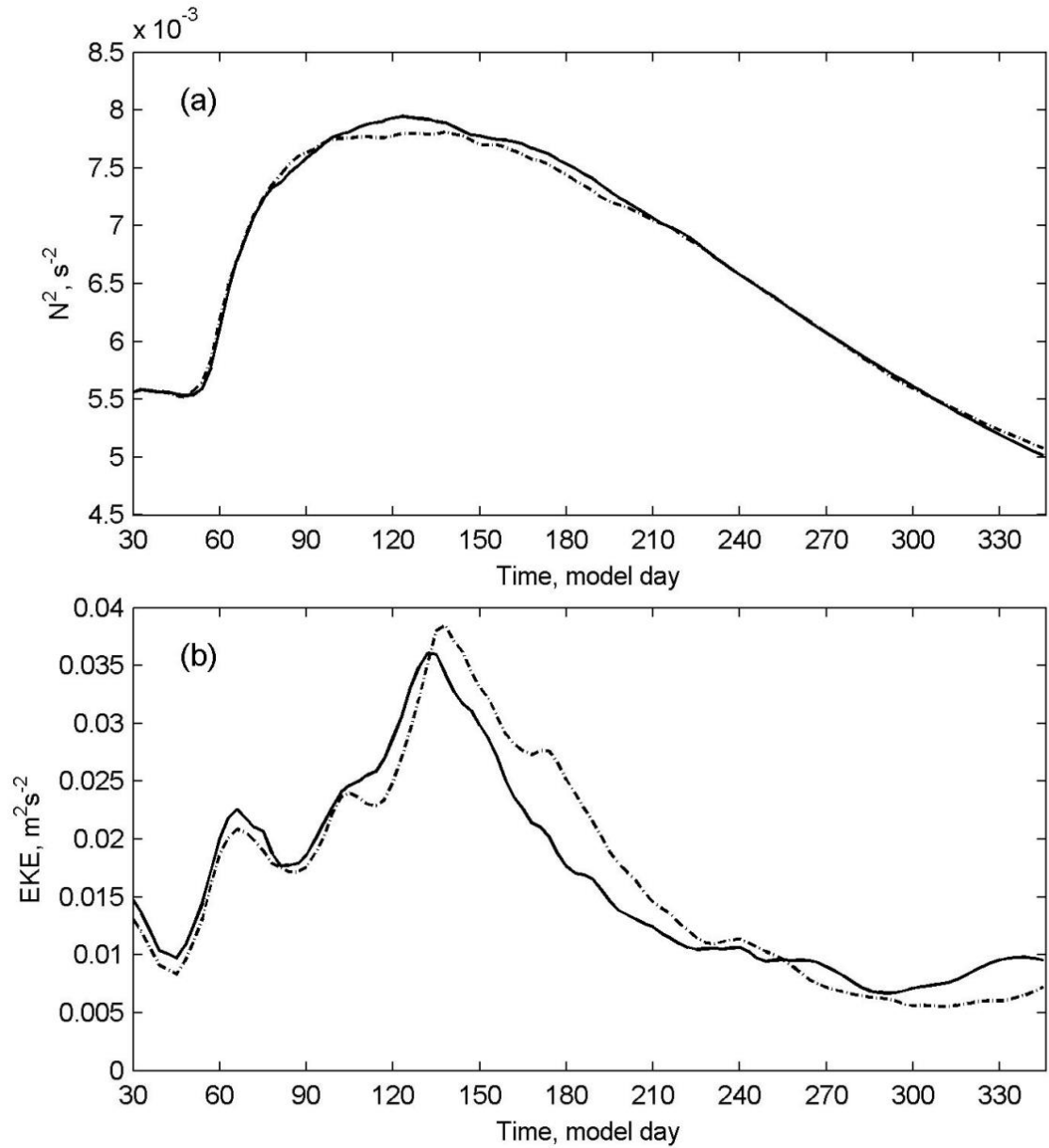

Figure 4. Time evolution of (a) buoyancy frequency ($N^2$) and (b) eddy kinetic energy (EKE) averaged over the horizontal plane at 50 m depth in (solid line) TCKD and (dot-dashed line) SCKD.

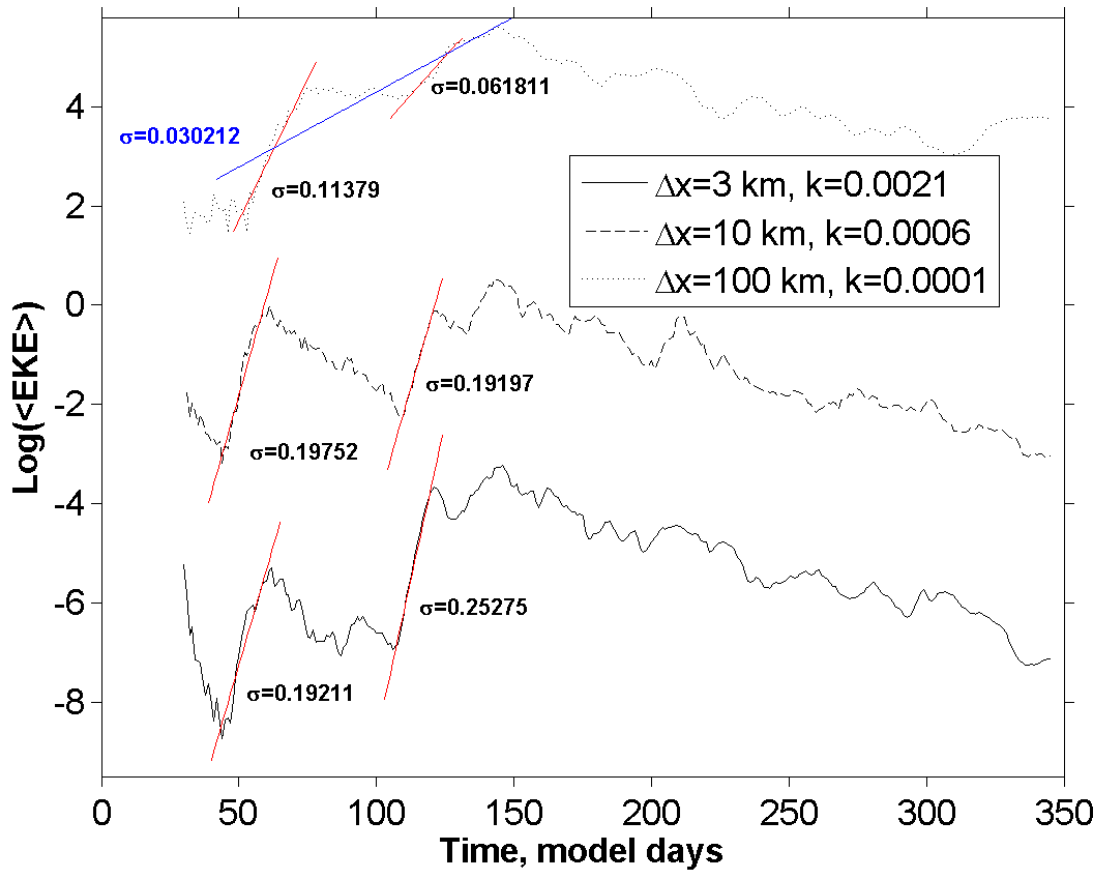

Figure 5. The natural logarithm of the eddy kinetic energy power spectra, <EKE>, in units of $m^4\ s^{-2}$ at three wave numbers listed in the legend, wave number k=0.0021 or wave length λ=3 km (solid line), k=0.0006 or λ=10 km (dashed line), and k=0.0001 or λ=100 km (dotted line). The red regressions lines identify the slope of <EKE> with time, which is the mixed layer eddy growth rate given by the σ value next to each red line. For each <EKE> line there are two periods for rapid eddy growth associated with the mixed layer instabilities. The blue line and associated growth rate is the deep baroclinic instability growth rate for the mesoscale eddies (k=0.0001 or λ=100 km).

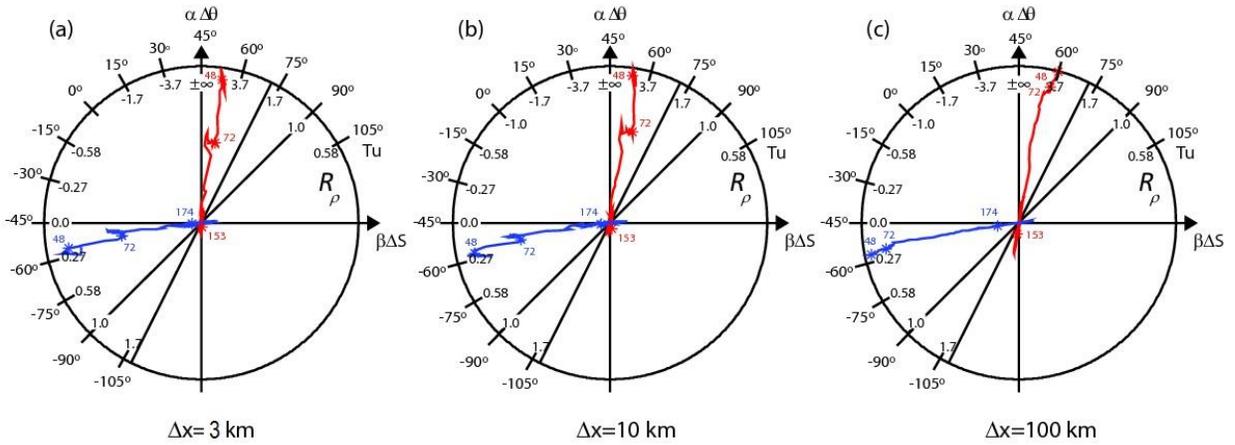

Figure 6. Time-evolution trajectories of density ratio $R_\rho$ (horizontal compensation angle, HCA) as functions of spatially-averaged cross-channel $\beta\Delta S$ and $\alpha\Delta\theta$ (eqn. 1) for spatial scales of (a) $\Delta x$=3 km, (b) $\Delta x$=10 km, and (c) $\Delta x$=100 km in (red curve) TCKD and (blue curve) SCKD at 50 m depth. The different dynamical stages of the simulation are marked with the stars (see Figure 3). The sectors of $R_\rho$=(1,2) representing conceptual transition of density compensation from the ML to the TC are shown with thin black lines.

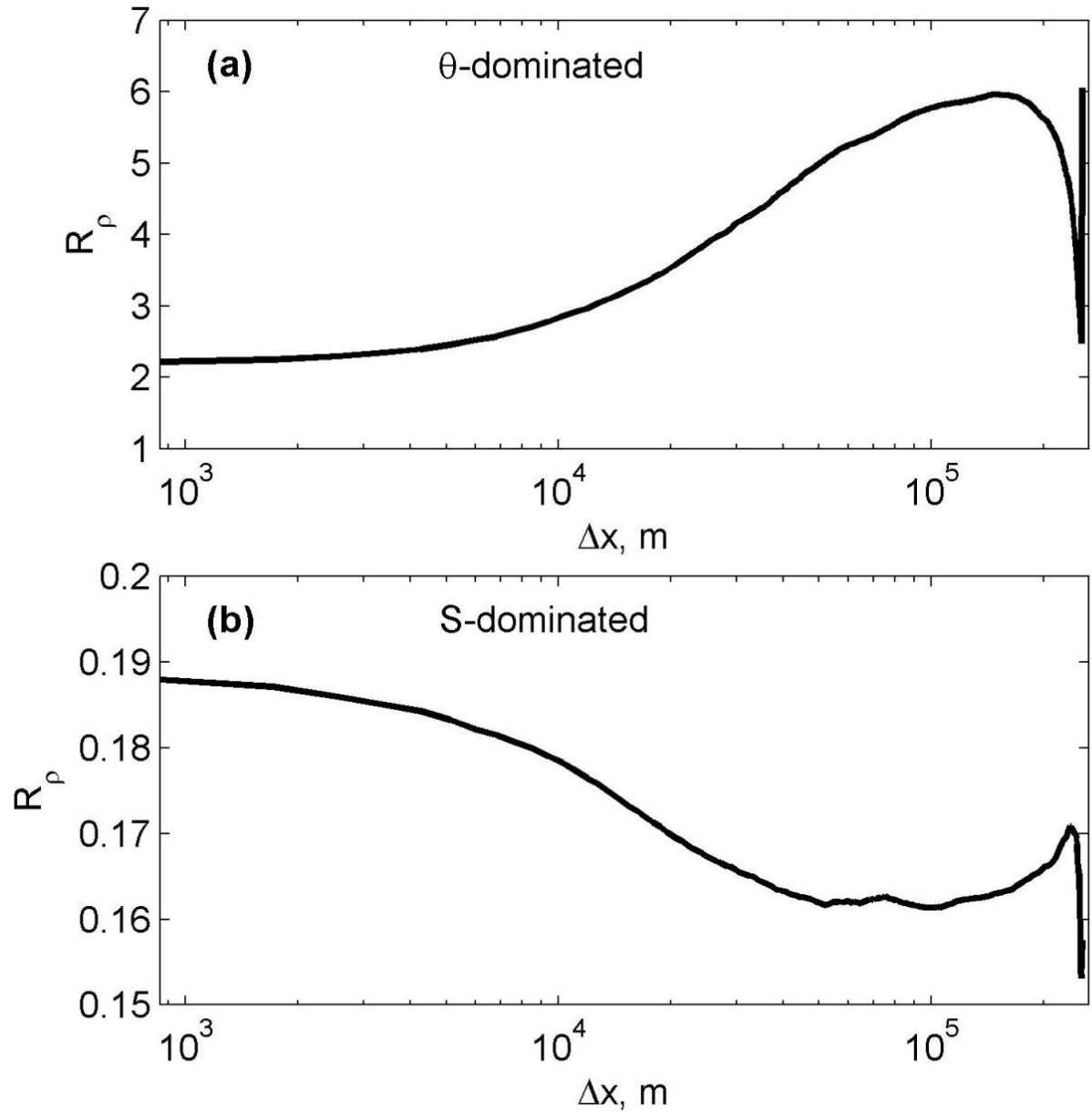

Figure 7. Spatially- and time-averaged horizontal cross-channel density ratio $R_\rho$ as a function of a spatial scale $\Delta x$ for (a) TCKD and (b) SCKD at 50 m depth.

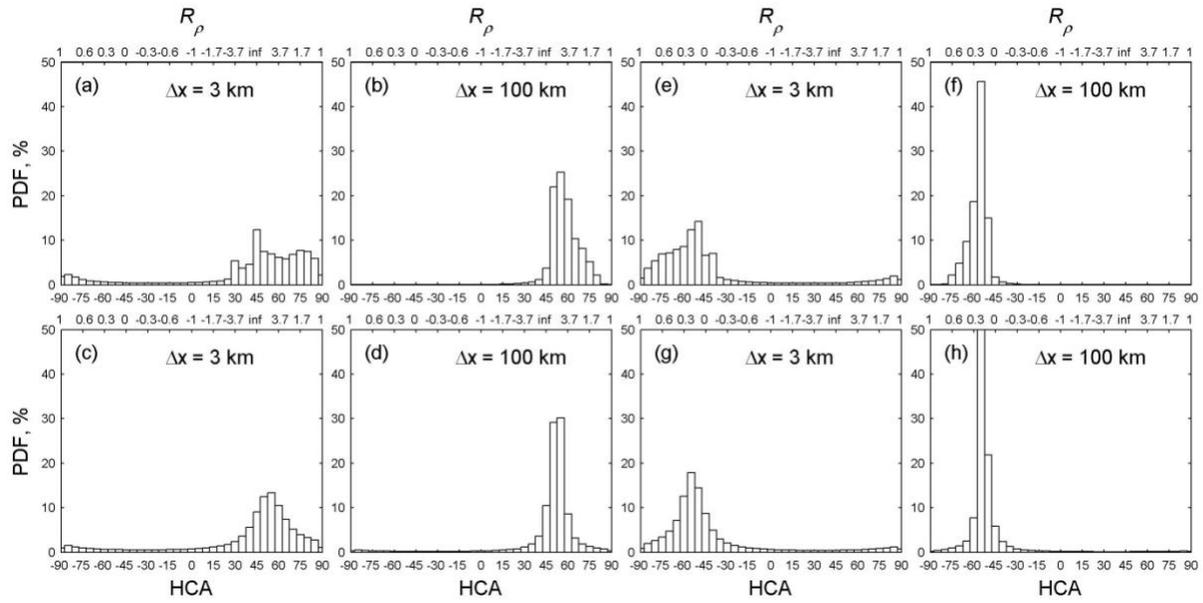

Figure 8. Probability density function (PDF, %) of cross-channel density ratio $R_\rho$ (horizontal compensation angle, HCA) for spatial scales of (a, c, e, g) $\Delta x$=3 km and (b, d, f, h) $\Delta x$=100 km in (a-d) TCKD and (e-h) SCKD at 50 m depth computed for the active mixed layer instabilty phase (a, b, e, f, model days 42 – 120) and for the rest of model integration (c, d, g, h, model days 123 – 345).

Page **29** of **33**

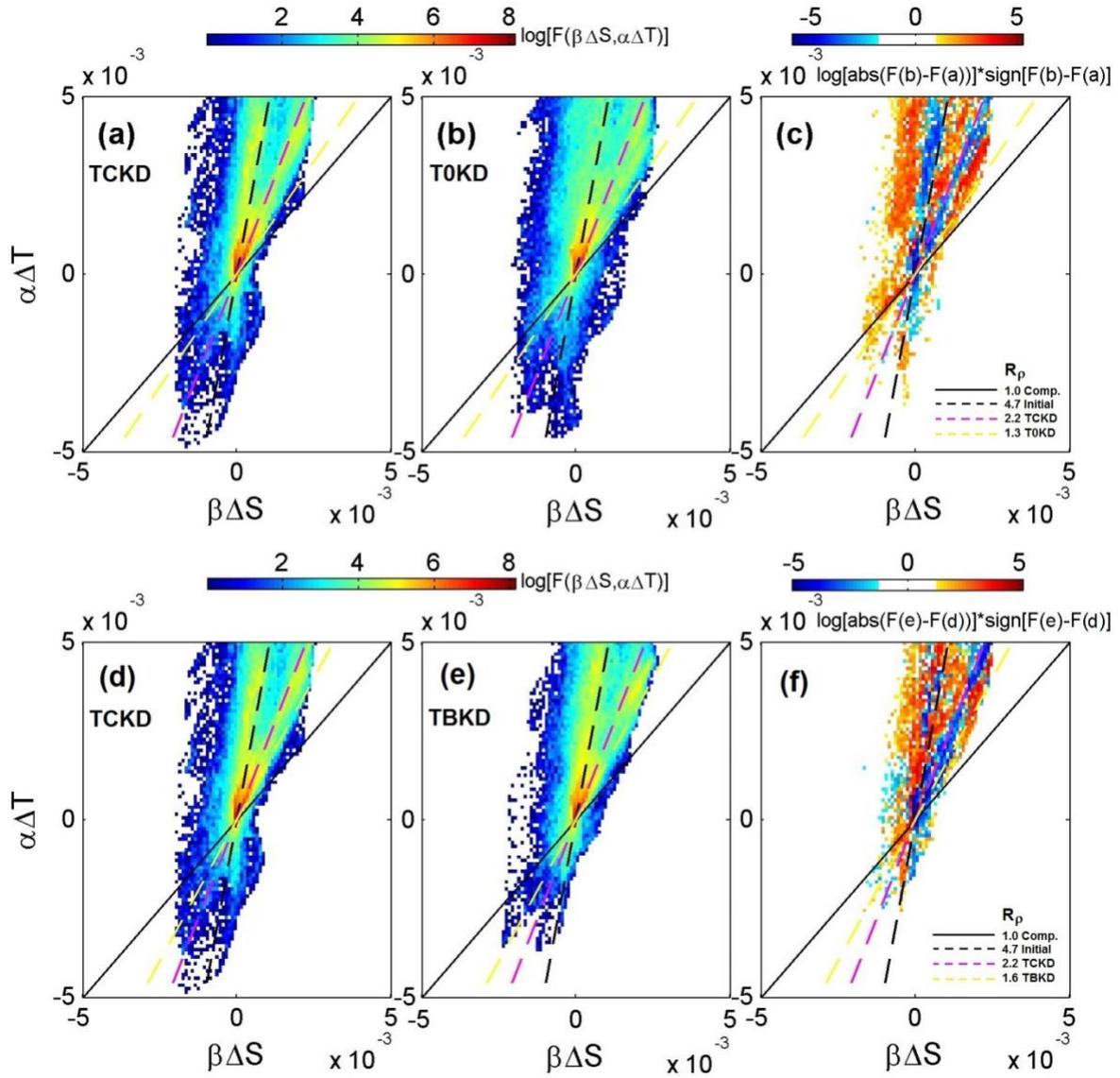

Figure 9. Natural logarithm of a function of joint distribution of $\beta\Delta S$ and $\alpha\Delta\theta$ computed for the range of spatial scales across the domain at 50 m depth for (a, d) TCKD, (b) T0KD, (e) TBKD (see Table 2), and a natural logarithm of the absolute value of the differences (c) T0KD-TCKD and (e) TBKD-TCKD multiplied by its difference's sign at model day 75 during the peak of active submesoscale circulation phase. Solid black line marks the compensation line with $R_\rho=1$; dashed black line corresponds to initial large-scale density ratio $R_\rho=4.7$ (Table 1); dashed magenta line – to TCKD small-scale ratio $R_\rho=2.2$; dashed yellow line – to (a, b, c) T0KD small-scale ratio $R_\rho=1.3$, and to (d, e, f) TBKD small-scale ratio $R_\rho=1.6$.

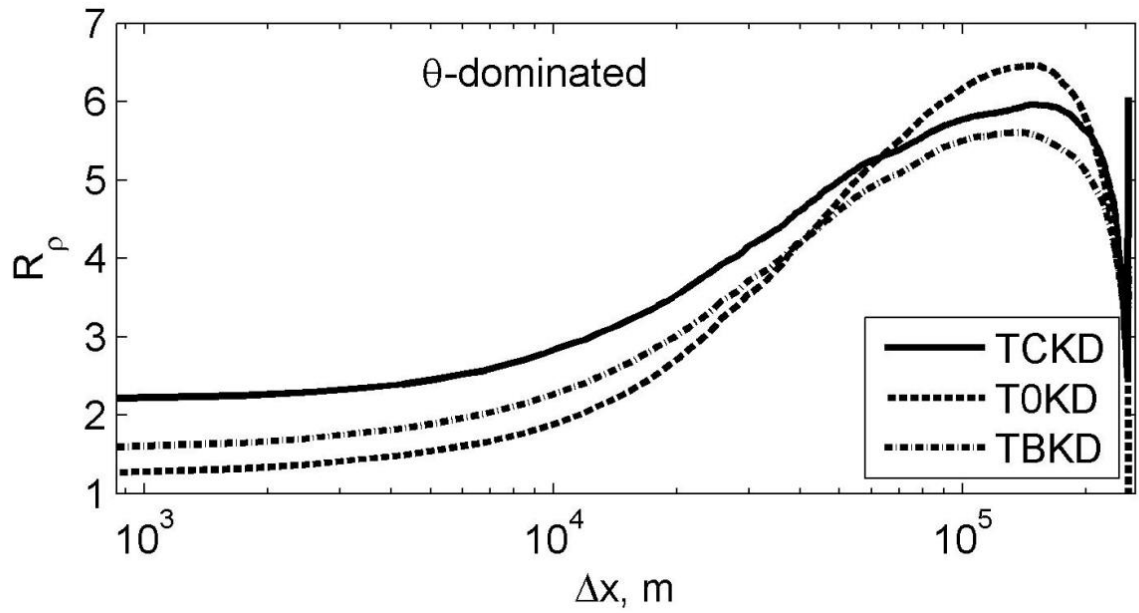

Figure 10. Spatially- and time-averaged horizontal cross-channel density ratio $R_\rho$ as a function of a spatial scale $\Delta x$ for temperature-dominated cases at 50 m depth for numerical experiments TCKD, T0KD, and TBKD with different horizontal diffusion parameterizations (see Table 2).

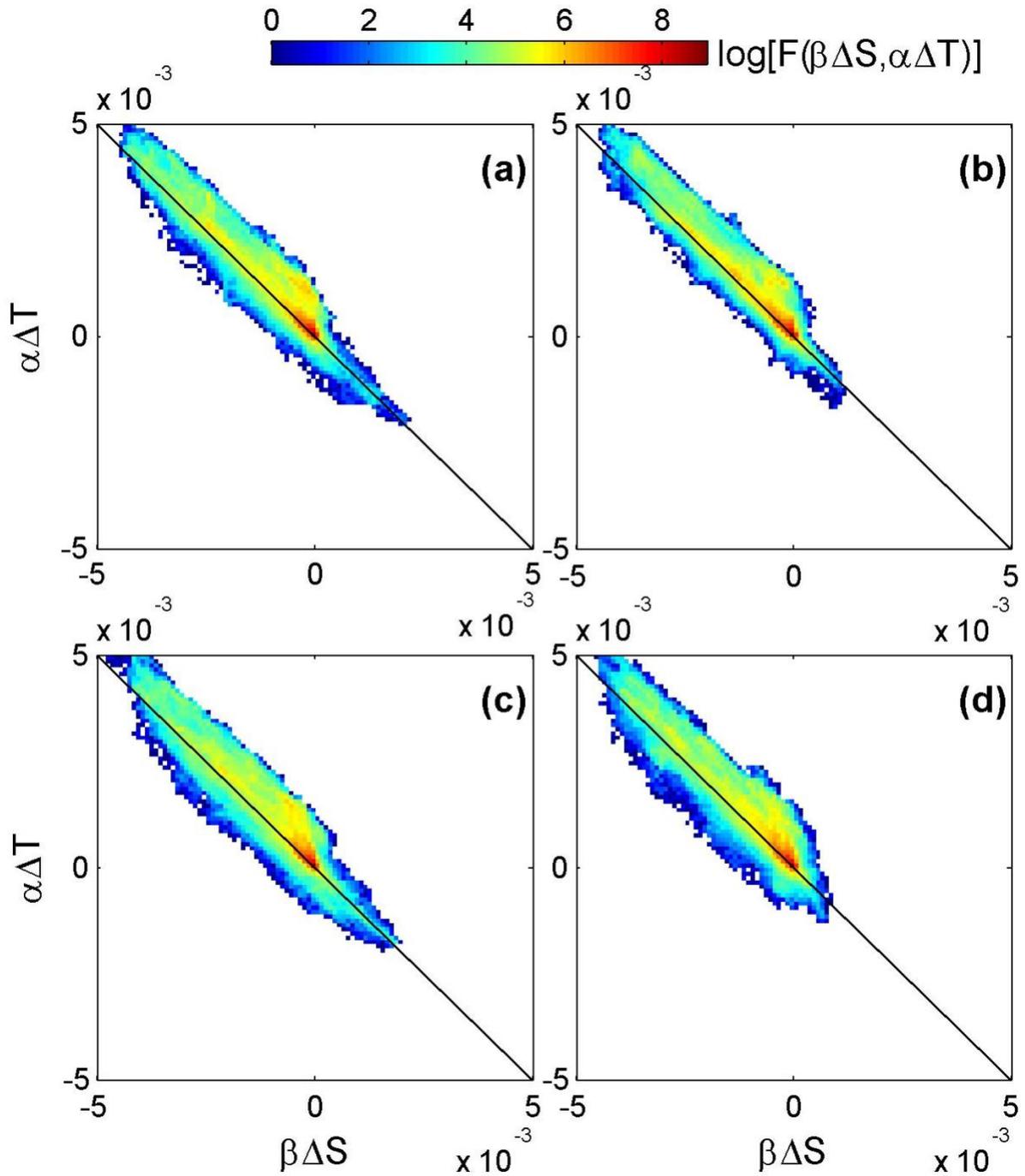

Figure 11. Natural logarithm of a function of joint distribution of *β∆S* and *α∆θ* computed for the range of spatial scales across the domain at 50 m depth for (a) NCKD, (b) NBKD, (c) NCB0, and (d) NBB0 (see Table 2) at model day 75 during the peak of active submesoscale circulation phase. Diagonal black line marks the cooperation line with $R_\rho$=-1.

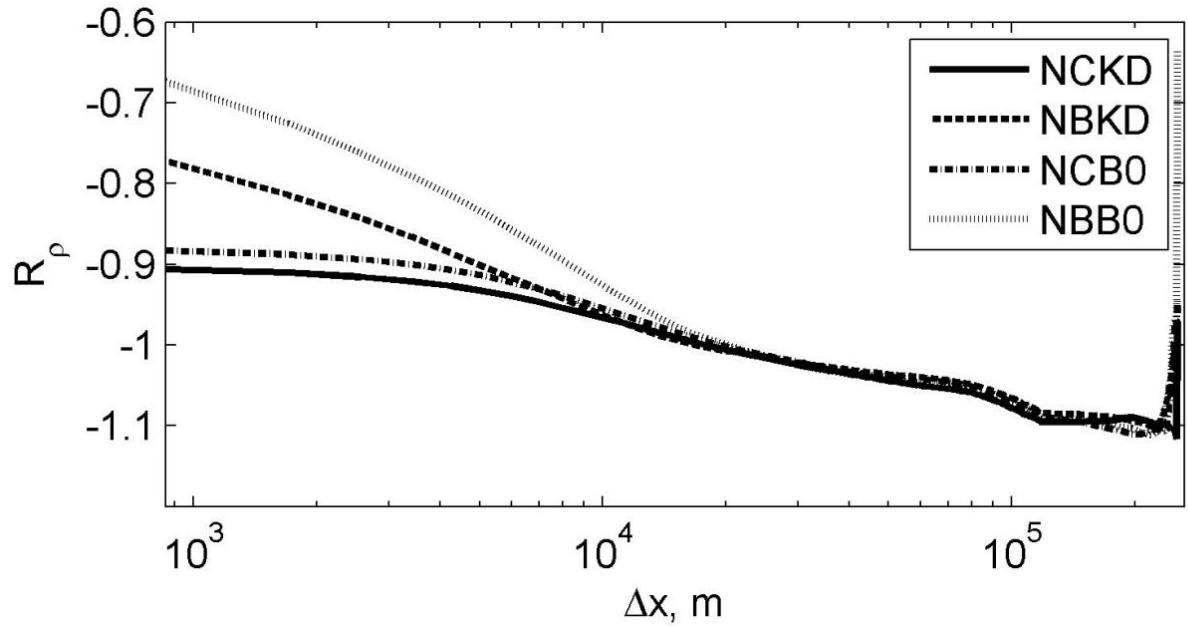

Figure 12. Spatially- and time-averaged horizontal cross-channel density ratio $R_\rho$ as a function of a spatial scale $\Delta x$ at 50 m depth for numerical experiments NCKD, NBKD, NCB0, and NBB0 with different horizontal and vertical diffusion parameterizations (see Table 2).